\documentclass[11pt]{article}

\usepackage{microtype} % microtypography
\usepackage{booktabs}  % tables
\usepackage{graphicx}
\usepackage{amsmath}
\usepackage{amsthm}
\usepackage{makecell}
\usepackage{url}
\usepackage{tabulary}
\usepackage{algorithm2e}
\usepackage{algpseudocode}
\usepackage{todonotes}
\usepackage{subcaption}

\usepackage[hidesupplement,final]{automl}

\title{Implementation of An Automated Learning System for Non-experts}
\author[1]{\nameemail{Phoenix X. Huang}{phoenix.huangx@gmail.com}}
\author[1]{\nameemail{Zhiwei Zhao}{thirteen2001@gmail.com}}
\author[1]{\nameemail{Chao Liu}{lc_kelvin@163.com}}
\author[1]{\nameemail{Jingyi Liu}{songlart@gmail.com}}
\author[1]{\nameemail{Wenze Hu}{windsor.hwu@gmail.com}}
\author[1]{\nameemail{Xiaoyu Wang}{fanghuaxue@gmail.com}}
\affil[1]{Lighthouse}

\hypersetup{%
  pdfauthor={}, % will be reset to "Anonymous" unless the "final" package option is given
  pdftitle={},
  pdfsubject={},
  pdfkeywords={}
}

\begin{document}

\maketitle
%\listoftodos
\begin{abstract}

Automated machine learning systems for non-experts could be critical for industries to adopt artificial intelligence to their own applications. This paper detailed the engineering system implementation of an automated machine learning system called YMIR, which completely relies on graphical interface to interact with users. After importing training/validation data into the system, a user without AI knowledge can label the data, train models, perform data mining and evaluation by simply clicking buttons. The paper described: 1) Open implementation of model training and inference through docker containers. 2) Implementation of task and resource management. 3) Integration of Labeling software. 4) Implementation of HCI (Human Computer Interaction) with a rebuilt collaborative development paradigm. We also provide subsequent case study on training models with the system. We hope this paper can facilitate the prosperity of our automated machine learning community from industry application perspective. The code of the system has already been released to GitHub\footnote{https://github.com/industryessentials/ymir}.
%{\bf Xiaoyu ran one pass}
% re-checked by Shiliang
\end{abstract} 

\section{Introduction}
Many AutoML techniques such as NAS and HPO have greatly simplified the task of finding the best models and hyper parameters for machine learning problems, and contribute to the wide adoption of ML in various fields. However, in industrial settings, modeling is only one of many factors which affect the efficiency and simplicity of a machine learning development process. A systematic approach that manages and automates the entire pipeline could be a key to land AutoML into in more application areas.

In ready to use AutoML systems such as Google Cloud AutoML, Amazon SageMaker Autopilot etc.\ , a model development pipeline is simplified to three steps: upload data, train model and deployment. While this simplified process achieves the goal of producing models, in actual use cases, the model development processes are almost always iterative, either because the raw data arrives in batches, or the labeling cost is too prohibitive that labeling has to be done in batches. In these settings, the time spent repeating non-training steps is usually much longer than training the models, rendering an automated pipeline more valuable than only automating the model training step itself.

YMIR\footnote{YMIR denotes You Mine In Recursive} is a machine learning platform that simplifies and automates the production grade machine learning processes in industrial settings. The system enables non-expert users to develop AI models through a no-code style GUI. To achieve this goal, YMIR defines its own generic model training and inference task protocols to support different algorithm implementations for various machine learning problems, its own task management and resource allocation framework to better organize and track machine learning tasks for multiple parallel projects, as well as its own GUI to emphasis an active learning based iterative machine learning development pipeline. Our case study in appendix shows that this platform can be used by non-experts to train and iterate models for production purposes, which helps wide adoption of AI to new application fields.

%Furthermore, experiments were conducted to show that the system is able to shift much of work load from algorithm developers to people without these expertise, which streamlines the development process for the ML engineers, and improves the efficiency of the entire model development process. Because of the work-load shift, YMIR enables elastic expansion of the dev-team's capabilities as it is possible to hire people without ML development experience to develop models. This helps the team to meet the demand surges encountered in rapid growth of a companies business.

It is noted that while the high-level concept design of YMIR is introduced in \cite{YMIR2021}, this paper emphasis on the engineering design principles as well as implementation details. The appendix includes case studies showing its effectiveness in speeding up the AI model development process.

\section{Industrial AI Model Development Process}\label{sec:inddevprocess}

YMIR is developed under the data-centric principle with focuses on efficient model and data iterations. As is shown in Figure \ref{fig:YMIR_process}, YMIR defines an iterative model development process as following steps:

\begin{figure}
\begin{subfigure}{\textwidth}
\resizebox{2.9in}{!}{
\SetKwComment{Comment}{/* }{ */}
\begin{algorithm}[H]
\KwData{$Data_{superset}$ \Comment*[r]{The whole available data}} 
\KwResult{Application Specific Model $M_o$}
$Acc_{T} \gets A$ \Comment*[r]{Target model accuracy}
$M_0 \gets$ Empty model or pretrained model \;
$D_0 \gets $ Empty, $Acc_0 \gets 0$, $i \gets 0$ \Comment*[r]{Initialize data, accuracy, iteration index}
\While{$Acc_i \leq  Acc_{T}$ \Comment*[r]{Continue if the model accuracy is not sufficient}}{
   
  \eIf{User Interrupts The Process \Comment*[r]{Terminated by user}}
  {
    $M_o \gets M_i$\;
    Break\;
  }{
    $d_i \gets$ Data mining output with $M_i$ and $Data_{superset}$ if $i > 0$, otherwise Initial Data \;
    Label $d_i$ \Comment*[r]{Label the mined data}
     $D_{i+1} \gets D_i \cup d_i $ \Comment*[r]{Update training data}
    $M_{i+1} \gets $ Model trained on $D_{i+1}$ \Comment*[r]{Update model}
    $Acc_{i+1} \gets $ Evaluate $M_{i+1}$ \Comment*[r]{Evaluate model}
  }
   $M_o \gets M_{i+1}$, $i \gets i+1$ \Comment*[r]{Update model}
}
\end{algorithm}
}
\end{subfigure}
\hspace{-3.1in}
\begin{subfigure}{0.5\textwidth}
\includegraphics[width=\textwidth]{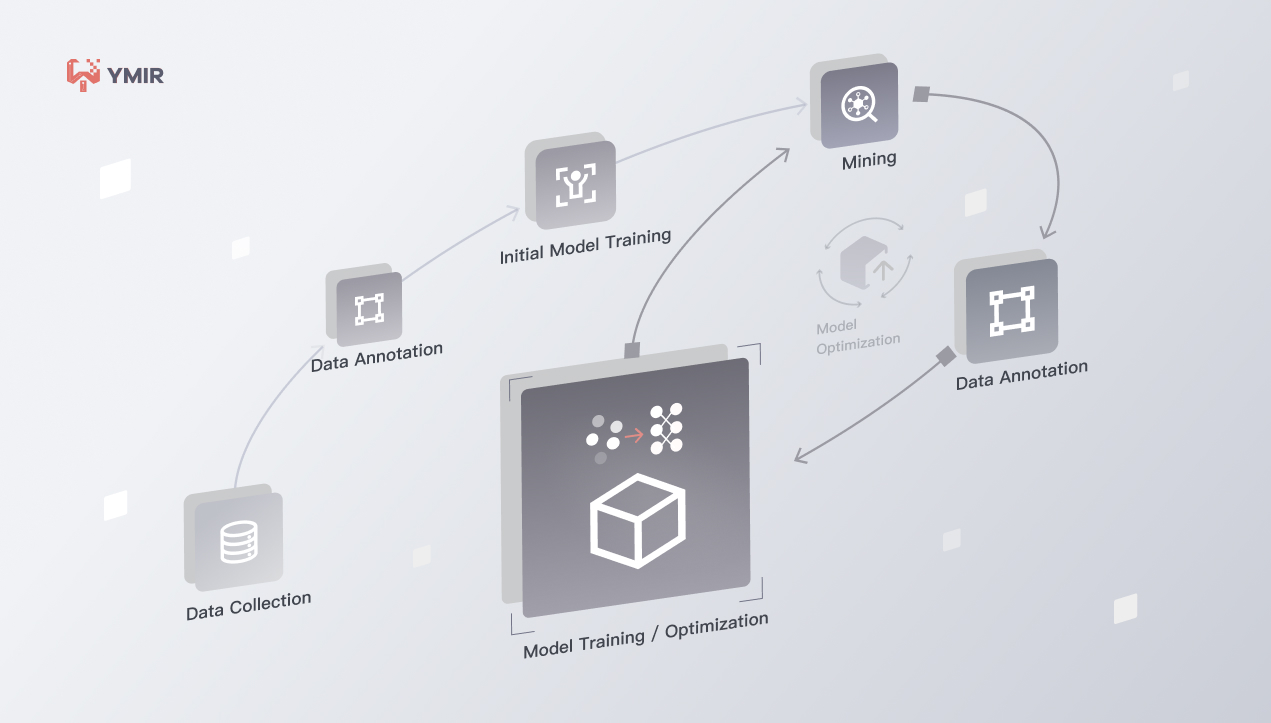}

\end{subfigure}
\caption{Model development process in YMIR system. Left: In algorithm format. Right: In the diagram format.}
\label{fig:YMIR_process}
\end{figure}

\begin{enumerate}
\item {\bf Solution Design: } This stage focuses on converting the business needs into an existing type of modeling problem. Taking computer vision as an example, the modeling problem could be image classification, object detection, segmentation, or a combination of multiple techniques. 

\item {\bf Data Preparation: } Data preparation mainly refers to the process of collecting and pre-processing data for a target technical solution. Data collection is usually performed iteratively until the resulting technical solution meets the application requirements.

\item {\bf Model Training}: Commonly known as "modeling", this is the step of exploring and analyzing prepared data through analytical tools, methods and techniques to discover input-output relationships and internal connections to inform decisions for business purposes. The result of this process is usually one or more machine learning models. The model training process is also iterative followed by data updates, until the model accuracy meets application requirements.

\item {\bf Model Evaluation}: This step evaluates performances of the models obtained from the model training step. Different from only using key metrics such as accuracy or sensitivity as diagnostic measures, which are usually computed on validation data while training the model, this process may involve more in-depth analysis of the results on testing datasets, such as model bias or metrics on sub-groups of the validation/testing sets.

\item {\bf Data Mining}: This stage focuses on finding the best data that improves the current models. On many applications this is a necessary step because the modeled events are usually rare events in the source dataset, and the cost of blindly labeling all the data could be prohibitive.

\item {\bf Model Deployment:} Trained and evaluated model needs to be applied to production data or newly acquired data to generate predictions which provide the high-value information for business purpose. The model usually is usually optimized in terms of computational cost in this step, to save the eventual cost of running them on real data.
\end{enumerate}

Processes $2$ through $5$ are repeated for improved model accuracy, while process $1$ and $6$ are separate procedures which are typically performed offline.  In following sections, we focus on the implementation of process $2$ through $5$. Figure \ref{fig:YMIR_process} illustrated the iteration logic in both algorithm format and graph format for better understanding.

\begin{figure}[ht]
\centering
\includegraphics[width=\textwidth]{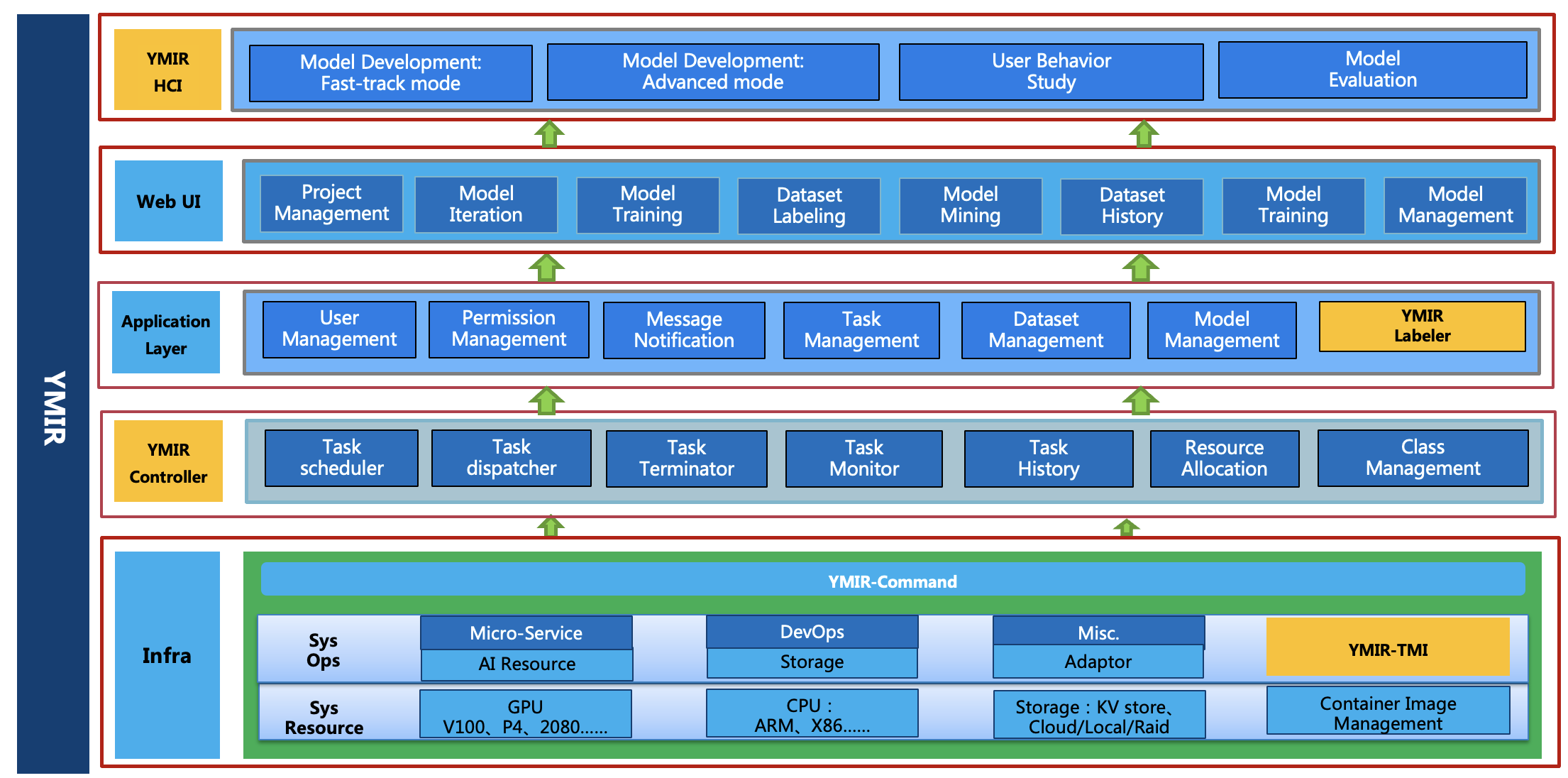}
\caption{YMIR decomposed into engineering layers. YMIR-HCI guides user operations and assists users to develop AI model more efficiently; YMIR-Web UI provides UI controls for users to visualize/operate dataset and tasks; YMIR-Application: this layer maintains the relationship among user/task/assets, as well as permissions; YMIR-Controller manages task framework and tracks the progress of individual task. The infra layer manages all infrastructure resources, data storage solution, \emph{etc}.}
\label{fig:YMIR_infra}
\end{figure}

\section{System Implementation}~\label{sec:design}
In YMIR, the pipeline introduced above is implemented through four key components, which are shown in Table  \ref{tab:YMIR_EngParts} and Figure \ref{fig:YMIR_infra}: 1) {\bf YMIR-TMI} implements model related operations including training, testing and data mining. 2) {\bf YMIR-Controller} implements task and resource management. 3) {\bf YMIR-Labeling} implements data labeling and connects to data labeling services through open interfaces. 4) {\bf YMIR-HCI} implements front-end user interface for the platform.

\begin{table}
\begin{center}
\resizebox{0.8\textwidth}{1.7in}{
    \begin{tabular}{p{1in}|p{5in}}
    \hline
        {\bf Modules} & {\bf Function Description} \\  \hline \\
       YMIR-TMI  &  Open design of model {\bf T}raining/{\bf M}ining/{\bf I}nference (TMI) services through docker containers. Adding a new TMI related algorithm can be achieved by replacing the corresponding docker image. \\  \\ \hline \\
        
       YMIR-Controller & Implementation of task and resource management. It decomposes ML tasks into atomic operations. Resource sharing, task progress monitoring and task orchestration are done by scheduling on top of these operations. \\ \\ \hline \\
       
       YMIR-Labeling & Open API for integration of data labeling softwares. \\  \\ \hline \\
       
       YMIR-HCI & Design of {\bf H}uman {\bf C}omputer {\bf I}nteraction (HCI). The graphical interface aims at implementing a rebuilt collaborative development paradigm, which guides users through the model and data iteration process. User actions as well as operation outputs together with key metrics are recorded, which makes the process trackable. The system improves the efficiency of data iteration and model production by interactive guidance through the whole process.\\ \\ \hline
    \end{tabular}
    }
    \end{center}
    \caption{Main Engineering Components of YMIR}
    \label{tab:YMIR_EngParts}
\end{table}

\subsection{YMIR-TMI}

\subsubsection{Design Principle}
\label{sec:tmi_design}

The focus of the YMIR platform is on data management and process control. When it comes to the  TMI services, we decouple these services from the platform and treat them as plug and play black boxes.  While YMIR provides default TMI dockers, users and third-party developers have the freedom to write and integrate their own related algorithms, as long as the corresponding docker images is compatible with the defined protocol. 

YMIR interacts with TMI services in the following way:
\begin{itemize}
    \item The YMIR platform, in conjunction with the TMI services, fulfills the whole development requirements.
a. the YMIR platform extracts the metadata for TMI services from docker images (descriptions,  configurations, etc.)
b. The YMIR platform provides input data to TMI services, and collects results when those service jobs are completed. 
c. The YMIR platform catches the exceptions and errors generated during the operation of TMI services.
d. The YMIR platform collects the real-time status of TMI services (whether a service is running normally,  its progress or real-time logs, \emph{etc}.)

\item A TMI service is decoupled from the YMIR platform. The service is responsible for the correctness on the implementation of corresponding model and data methodologies.

\item The YMIR platform is not aware of the internal dependencies of these services.

\item There is no dependency between the YMIR platform and the versions of these services.
\end{itemize}

\subsubsection{Implementation}

\begin{table}[ht]
\begin{center}
\resizebox{\textwidth}{!}{
\begin{tabulary}{\textwidth}{ p{.5in} p{1in}LL }
 Name & Interaction & Flow Direction & Notes \\ 
 \hline
 Metadata & When the TMI service is registered. & YMIR reads from TMI services. & A description of the TMI service itself, and the list of key configurable parameters. \\
 Logs & During the operation of the TMI instance. & Generated by the TMI instance, read and logged by YMIR. &  The runtime logs generated by the TMI instance are only stored by YMIR and do not need to be understood. \\
Status Monitoring & During the runtime of a TMI instance. & Generated by the TMI instance (written to a location-specific file using a defined format), read by YMIR & Monitoring information generated by the TMI instance about its latest status. \\ 
 Input & Before the TMI instance runs. & Generated by YMIR (placed in a specific path and mounted at boot time) and read by the TMI instance & For the training service: training set, test set, pre-trained model, service parameter profile;  For mining service: model, candidate set, service parameter profile; For the inference service: model, set of images to be inferred, service parameter profile. \\ 
 Output & After the TMI instance is completed & Generated by the TMI instance (written to a specific location using a defined format) and read by YMIR. & For the training service: model, result parameters file;  For mining service: mining result files; For inference service: inference result files.
\end{tabulary}}
\caption{Summary of the data exchange between YMIR system and TMI services.}
\label{table:data_exchange}
\end{center}
\end{table}

The TMI services are encapsulated into separated docker images, and the YMIR platform follows the life-cycle described in Section \ref{sec:inddevprocess} to invoke these services. YMIR instantiates TMI services and interact with them following principles described in \ref{sec:tmi_design}. In the following we explain TMI service registration and the service runtime.

\begin{enumerate}
\item \textbf{Service registration}. A service is available only after it is registered in YMIR. The availability ends when the service is logged off. For a registered service, the user should make sure:

\begin{enumerate}
\item The YMIR platform is aware of the metadata associated with this service.
\item The docker image which implements the specific task is ready to be instantiated.
\end{enumerate}

\item \textbf{Service runtime}: The system can instantiate one or more services after the service is registered. Unless exited with errors, the respective service keeps in the runtime until the job is finished. TMI instances in runtime period have the following characteristics.

\begin{enumerate}
\item Different TMI instances do not interfere with each other. There is no communication between instances, either.
\item YMIR has real-time access to the operational status of these instances.
\item YMIR is notified when a instance finishes running or exits with errors.
\end{enumerate}
\end{enumerate}

Table \ref{table:data_exchange} summarizes data exchange between the YMIR platform and TMI services instances.

\begin{figure}[ht]
\centering
\includegraphics[width=0.7\textwidth]{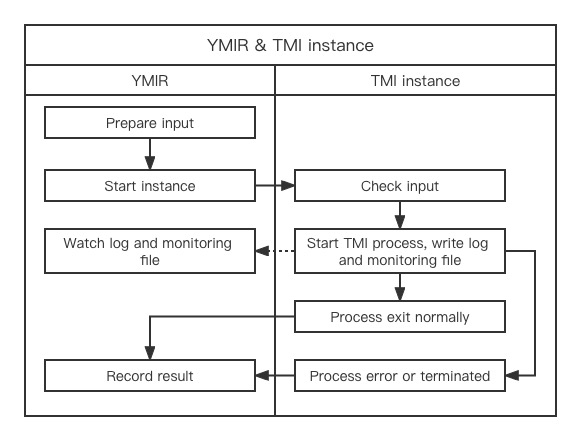}
\caption{Overview of the TMI service life cycle.}
\label{fig:tmi_lifecycle}
\end{figure}

The life cycle of TMI services is illustrated in Fig~\ref{fig:tmi_lifecycle}. It consists of multiple stages.

\textbf{Instantiation start-up.} YMIR follows the listed steps below to start a TMI service:

\begin{enumerate}
    \item The user chooses a service (docker image) to start with, and sets up necessary configurations for the service.
    \begin{enumerate}
        \item For the task of model training, the configuration of training sets and testing sets is mandatory. YMIR allows users to configure their pre-trained models for initialization if there is any.
        \item For the task of data mining or inference, only the target mining dataset or the dataset for inference is needed. 
    \end{enumerate}
\item YMIR platform reads assets configured in Step 1 and hosts them in predefined locations.

\item The TMI instance (docker container) starts its main process, taking the datasets, models and configuration files from predefined locations.

\end{enumerate}

\textbf{Instance status monitoring.} When the main process of the instance starts, the YMIR platform interacts with this instance in two ways.
1. All standard and error outputs generated by the instance are logged to a file.
2. The latest status and progress of the instance is written to a status monitoring file, and changes to this file are logged by the YMIR platform.

\textbf{End of instance.} If the instance executes all operations and returns a true value, the YMIR platform will consider it as a successful run. For these instances, the YMIR platform stores the results and updates the status of the tasks. If the instance exited unexpectedly or the return value is not true, the YMIR platform marks the task as a failure and archives the intermediate results (if any). If the user chooses to manually stop this instance during the execution of an instance, the YMIR platform sends a docker stop command to this instance, collects any intermediate results that may exist, and marks this task as broken.

\subsection{YMIR-Controller: Implementation of Task and Resource Management}

In YMIR, machine learning tasks are decomposed into smaller granular atomic-level tasks, which can be reorganized to achieve reuse of resources, progress monitoring at a smaller granularity. Task parallelization also benefits from the design.

\subsubsection{Task Scheduling}
The task scheduling component manages to connect a series of subtasks to implement the task defined by the user. It supports tasks parallelization if there are enough resources to support these tasks.

% \begin{figure}[h]
% \includegraphics[width=0.6\textwidth]{figure/arch4.png}
% \end{figure}

\begin{enumerate}
\item Data preparation subtask. The dataset preparation stage is separated, and only one copy of this original data of the image is kept, and the original content of the image is calculated using sha256() as the identifier of the image, which is reused in other tasks to avoid multiple copies of the original data, taking up storage space and task time, etc.
\item Dataset processing. Filtering, merging, taking intersection and merging of datasets, etc., will generate a new dataset after completion, which can also be reused by other tasks
\item GPU resource allocation. When the data is processed, before using GPU resources for training, the GPU ID used is stored from the task layer perspective, and the GPU resources are released when the task is completed or failed. Thus, the GPU resources are allocated in an on-demand fashion.
\item TMI execution. TMI services run respective jobs given all related resources are ready . %\todo{Update TMI execution}.
\end{enumerate}
\subsubsection{Assets Replica Management}

The target speed of data caching layer for data acquisition is within 100ms. It is organized as:
\begin{enumerate}
\item Original file parsing and verification. Get the hash\textunderscore id of the image content, and support reading the corresponding PASCAL, VOC, YOLO and other common data formats, where the hash\textunderscore id is used for image de-duplication and subsequent identification of the image. At the same time, it can support different strategies to ignore unknown tags, abort unknown tags, etc. to provide more information for later data analysis.
\item Extract key information, compress and store. After the data is obtained, it will be compressed and serialized using Protocol Buffer, and git will be called to store it, thus compressing the storage disk space, and also allowing to retrace the version record information.
\item Cache the data. The hash \textunderscore id is stored in an array, while other information such as image annotation is placed in a chain table with the head pointer of the chain table as hash \textunderscore id. hash \textunderscore id represents both the image itself and the nodes of the chain table, the array can be used for fast paging, and the chain table ensures O(1) to find the details of the image. Since the check phase has been de-duplicated in the previous stage, there is no problem of conflict
\end{enumerate}

\begin{figure}
\centering
\begin{subfigure}{0.28\textwidth}
    \label{fig:YMIR_hash_node}
    \includegraphics[width=\textwidth]{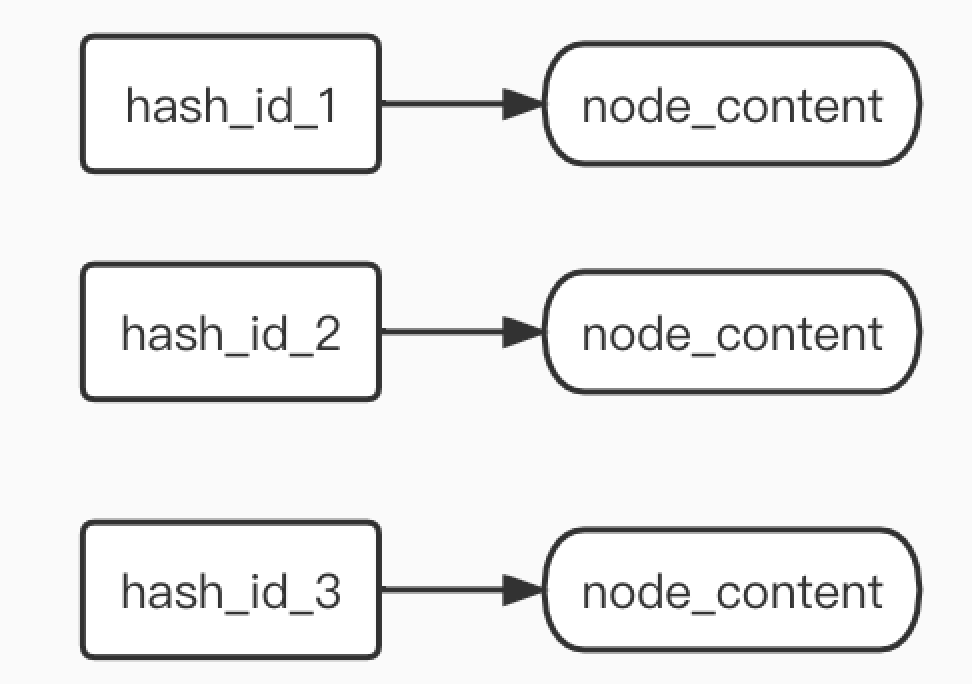}
    \caption{Hash ID}
\end{subfigure}
\begin{subfigure}{0.7\textwidth}
%\begin{figure}[h]
%\centering
    \label{fig:YMIR_cache}
    \includegraphics[width=\textwidth]{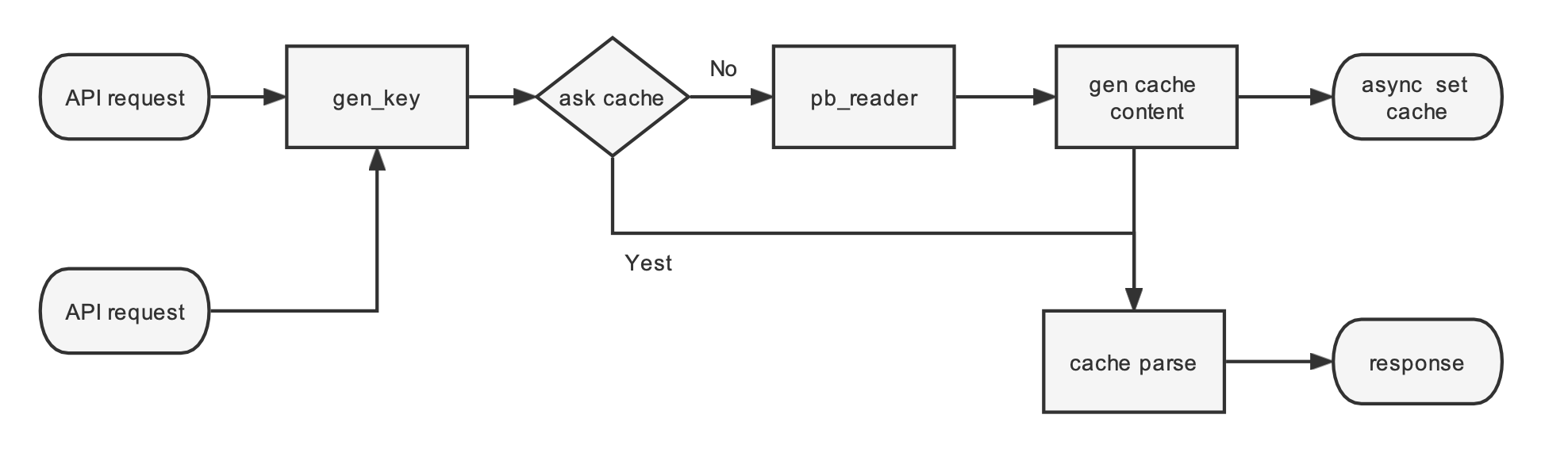}
    \caption{Caching}
\end{subfigure}
\caption{(a):Each YMIR asset has a unique ID. (b) Workflow of YMIR caching.}
\end{figure}

\subsubsection{Task progress tracking}
Background
The process of synchronizing task progress information through command, application and web UI layers is implemented in actively push mechanism. The logic of updating task progress is as follows.
\begin{enumerate}
\item YMIR-TMI and YMIR-Command write task progress to monitor.txt
\item YMIR-Monitor component polls task progress via task\textunderscore monitor, and posts to application layer.
\item YMIR-Application layer receives task progress message, and pushes the progress to web UI through the protocol tunnel, and displays it on the interface.
\end{enumerate}

\textbf{Functional design and implementation}
\begin{enumerate}
\item YMIR-Monitor collector in controller layer is responsible for collecting task progress information (user-id, task-id, percent, state code, state message, error message, timestamp, etc.) from monitor.txt
\item YMIR-Monitor collector sends the signal of task progress that has changed since last updated time. The dispatcher service at the app level posts this signal to a dedicated API at a certain interval. This signal is formed by the task id, state code, percent, etc.
\item web clients connect to the name-space of the current user via socketio.
\end{enumerate}

To post a progress message, the event dispatcher:

\begin{enumerate}

\item Create a message subscription group x for the Redis stream, with only one consumer

\item Create an API interface to the controller to receive information about tasks that have changed in the meantime, and put them into the Redis stream queue as a list when they are received

\item When the message subscription group x detects a change in the queue, it reads all messages from it at once, groups them by user-id and task-id, finds the latest task status corresponding to each task-id, pushes it to the namespace corresponding to the user-id using socketio, and writes the latest task status to the db via the API. If the write process fails, the categorized messages are rewritten back to the queue, and if the write process is successful, the original messages are deleted.
\end{enumerate}

Web UI subscribes to the event dispatcher service for individual user messages (connected to the namespace corresponding to the user-id) via socketio. The received messages are displayed to web users.

\subsection{YMIR-Labeler: Integration of Labeling Software}

Similar to YMIR-TMI system, the YMIR-Labeler component is also open-designed to support easily integration of external labeling tools. The whole life-circle of a labeling task is described in Figure \ref{fig:YMIR_labeler}. It includes following functionalities:

\begin{figure}[ht]
\centering
\includegraphics[width=0.9\textwidth]{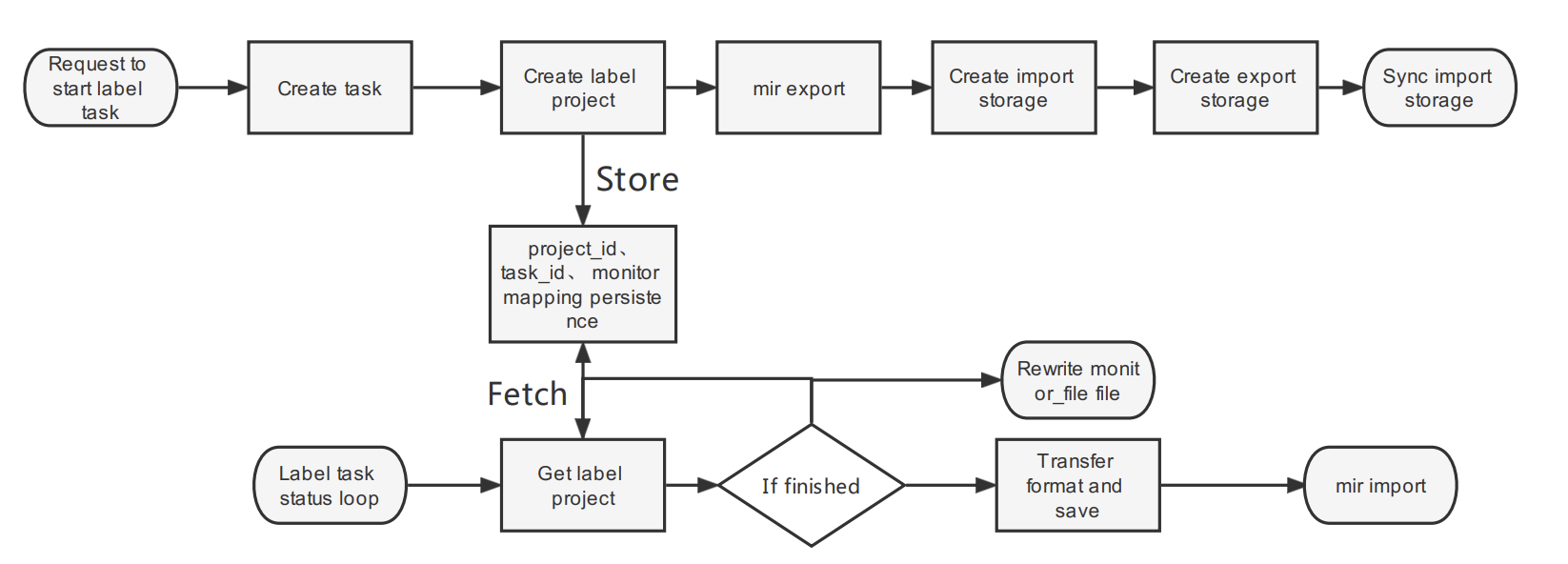}
\caption{YMIR labeling task life circle.}
\label{fig:YMIR_labeler}
\end{figure}

\textbf{Functionalities}

\begin{itemize}
\item Data labeling support for various vision tasks such as classification/detection/segmentation, \emph{etc}.
\item Pre-annotation using a pre-trained model to improve labeling efficiency.
\item Dataset synchronization between the labeling system and YMIR.
\item YMIR is able to distribute tasks and retrieval results from the labeling platform.
\end{itemize}

%\textbf{Labeler implementation}

The open design allows YMIR to use different labeling systems which may have their own merits for a specific task. As long as the defined communication interface is implemented, a third party labeling platform can be used by YMIR in a plug-and-play fashion.  

With an unlabeled dataset already in YMIR, the labeling task can be simply activated by clicking the {\bf [New Annotation Task]} button in YMIR's task management page. Users will be redirected to the annotation task creation page to input necessary labeling configurations and instructions. Once a labeling task is successfully created, one can check the corresponding progress by simply visit YMIR's task management page. YMIR is implemented to keep communicating with the labeling software and collect all the annotation results once the task is completed.  

\subsection{YMIR-HCI: Iteration Pipeline Centered Design.}

\begin{figure}[ht]
\centering
\includegraphics[width=0.95\textwidth]{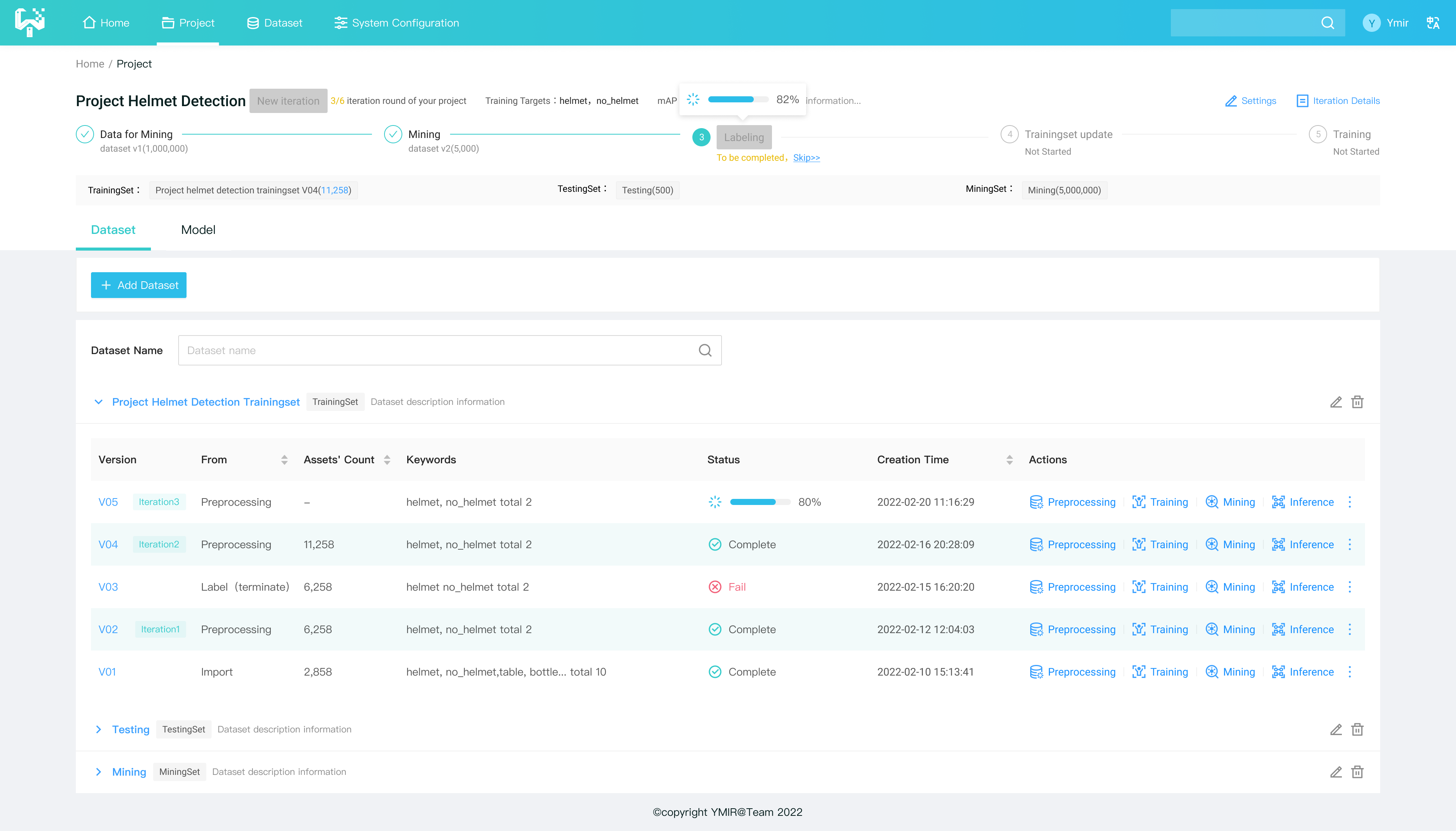}
\caption{YMIR System: Guided Interactive Model Development Pipeline. }
\label{fig:projects_view}
\end{figure}

Figure \ref{fig:projects_view} shows the screenshot of the project page for model development. A pipeline composed of connected buttons resembling an actual model iteration process defined in Figure \ref{fig:YMIR_process} (a) is put on the most prominent area of this page (see section "Project Helmet Detection" which includes connected actionable items such as labeling, training, data mining, \emph{etc}.

This pipeline gives users a clear view of the steps needed for performing one model iteration, progress of the current iteration and overall status of the running step. While providing high level information and actionable links to more details, this visual pipeline gives user the implication that model iteration is an important aspect of actual ML development project. This impression also helps inexperienced users realize that in application-oriented ML projects, developing a high quality dataset is equally if not more important than changing the model architectures or tuning the model training hyper parameters.

After a step is finished successfully, YMIR will guide the user to start the next step by highlighting the button automatically, following the logic specified in Figure \ref{fig:YMIR_process}.  The next step can also be configured to get triggered automatically, though we notice that many users prefer to inspect the results before proceeding to the next step.

YMIR also provides a project view page, to facilitate users develop multiple ML projects in parallel. More screenshots of YMIR UI are listed in Appendix.

\section{Limitation and Broader Impact}~\label{sec:LB_statement}
The system only supports training vision models. As a model production application, it should not have negative social or ethic impacts.

\section{Conclusions}~\label{sec:conclusion}
This paper introduces the engineering implementations of a machine learning system that uses a rebuilt development pipeline to involve non-experts into ML development process. We hope this project can facilitate the prosperity of our automated machine learning community from an industry application perspective.

\bibliography{egbib}

\newpage

\section{Appendix I: Case study, Developing AI models without AI Experts} 
This appendix presents a case study conducted with a ML modeling team that uses YMIR to improve their efficiency on solving real world problems.

For a time span of 2 weeks, a group of 1 ML engineer and 2 data annotators teamed up to try iterating models using YMIR. Before using YMIR, the ML engineer usually iterates 2 models a month, each with 2 iterations. Now, the 3 people group iterated 4 models only using two weeks. On average, the ML engineer reported significantly less time spent on iterating these models, from 0.3 month per model to 0.05 month. The ML engineer got extra 'free' time by using YMIR, which were spent on developing new algorithms and planning on new model development projects.

%\includegraphics[width=\textwidth]{figure/arch9.png}
%From Jan. 10 to Jan. 25, 2 annotators iterated 4 long-tail algorithm models orgranized by 1 researcher, the following is the case of iteration, compared with the manpower cost of 0.3 researcher-month to develop a long-tail models, after using YMIR, some tedious and repetitive operations can be operated by annotators in a systematic way with mouse click, and the development cost of a single algorithm can be reduced to 0.05 researcher-month + 0.4 annotator-month.

\begin{figure}[h]
\includegraphics[width=\textwidth]{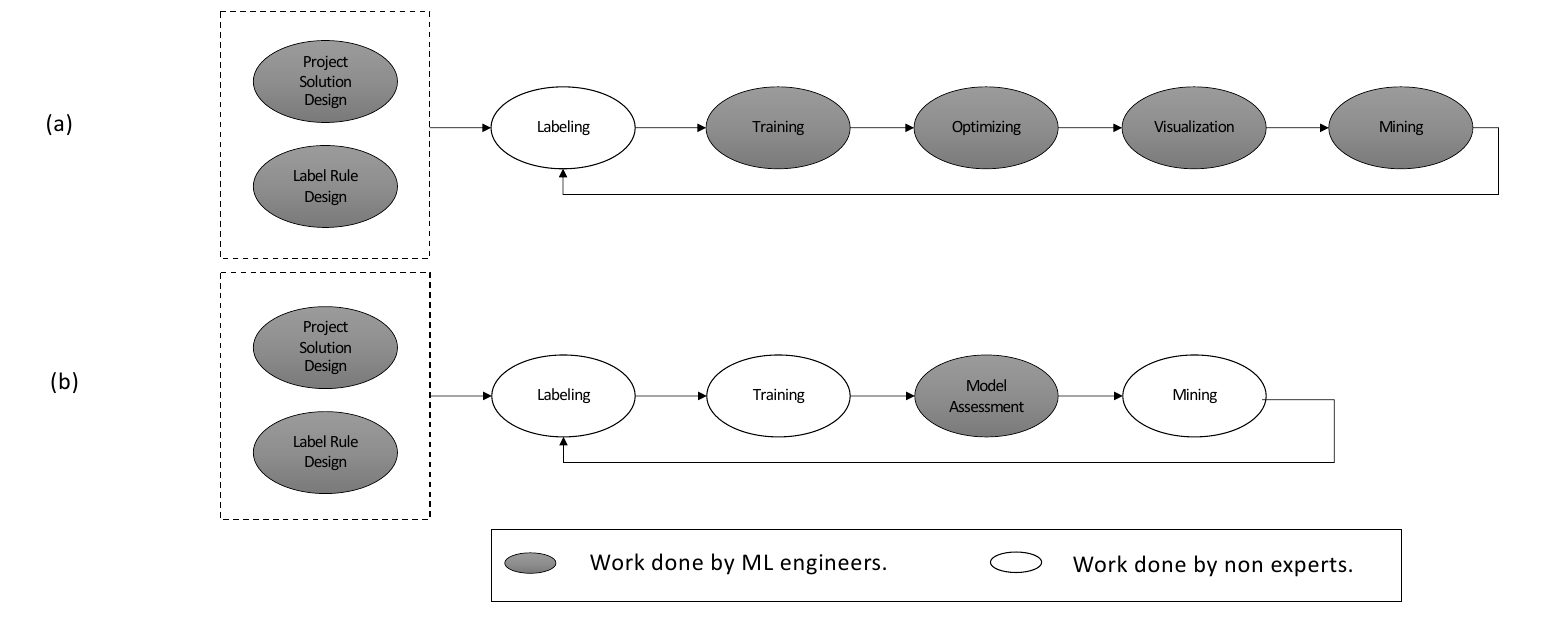}
\caption{(a) Original ML dev pipeline. Most of the steps are done by ML engineers. (b) New dev pipeline with YMIR, which enables non-experts to develop models.}
\label{fig:process}
\end{figure}

Comparison and analysis of different model development processes.

Figure \ref{fig:process}(a) shows the regular process of algorithm development by professionals aka ML engineers. The main operations of these engineers are model training, visualization, optimization, and mining, all of which needs to be done by using code. After the mining step, the selected data is handed over to the labeling group for labeling, and the training starts again after the mined dataset is labeled.

\textbf{Disadvantages of traditional ML development workflow.}

\begin{itemize}
\item Switching between different algorithm tasks usually takes time, and when the data is fully labeled, the algorithm staff has other work at hand to finish, which makes the prepared data idle waiting for engineer's availability.
\item Switching between different operations requires code implementation, and the efficiency of development depends on the coding level of the specific engineer.
\item The algorithm development process cannot be tracked. If many algorithms are developed in parallel, the algorithm developer needs to manually check the status of each task. Depending on the developer's work style, this could be a major loss of productivity that can be boosted by automatic tools.
\end{itemize}

\textbf{Advantages of traditional ML development work flow.}
\begin{itemize}
\item High flexibility in algorithm development, allowing selection of the most appropriate algorithm model based on project requirements.
\item Multiple tools can be used to tune the algorithm, such as adjusting the model training parameters, visualizing and analyzing the model output, and adjusting the training and labeling strategies based on the model visualization results.
\item Use analysis tools to monitor model training curves, identify problems early and correct them. This may save computational resources, and avoid wasting time on problematic datasets to produce low performing models.
\end{itemize}

Figure \ref{fig:process}(b) shows the model development process with YMIR. It can be observed that the majority of the operations are performed by the annotators, and the algorithm staff only reviews the model after it is trained and then moves his/her time on to the next task.
    
\textbf{Advantages of using YMIR for development}
\begin{itemize}
\item Switching between tasks and operations is simple and can be done with a single mouse click.
\item Data that has been annotated can immediately be used in the next step without waiting for the ML engineer to start the job manually.
\item One operator can develop multiple algorithms at the same time.
\item Reduce the knowledge requirement of algorithm development, so that people with no development experience can train and iterate models according to the operation documents.
\item Avoid tedious code operation switching, and development efficiency is less sensitive to the technical capability of the developer.
\item The entire model iteration process is recorded and trackable through the platform. The iteration history can be useful for further investigation of model related issues, such as analyzing the source of bias and fairness problems. It can also be used to spot, share and promote best practices when dealing similar tasks.
\end{itemize}

\section{Appendix II: Screenshots of YMIR system UI} 

\begin{figure}[h]
\centering
\includegraphics[width=\textwidth]{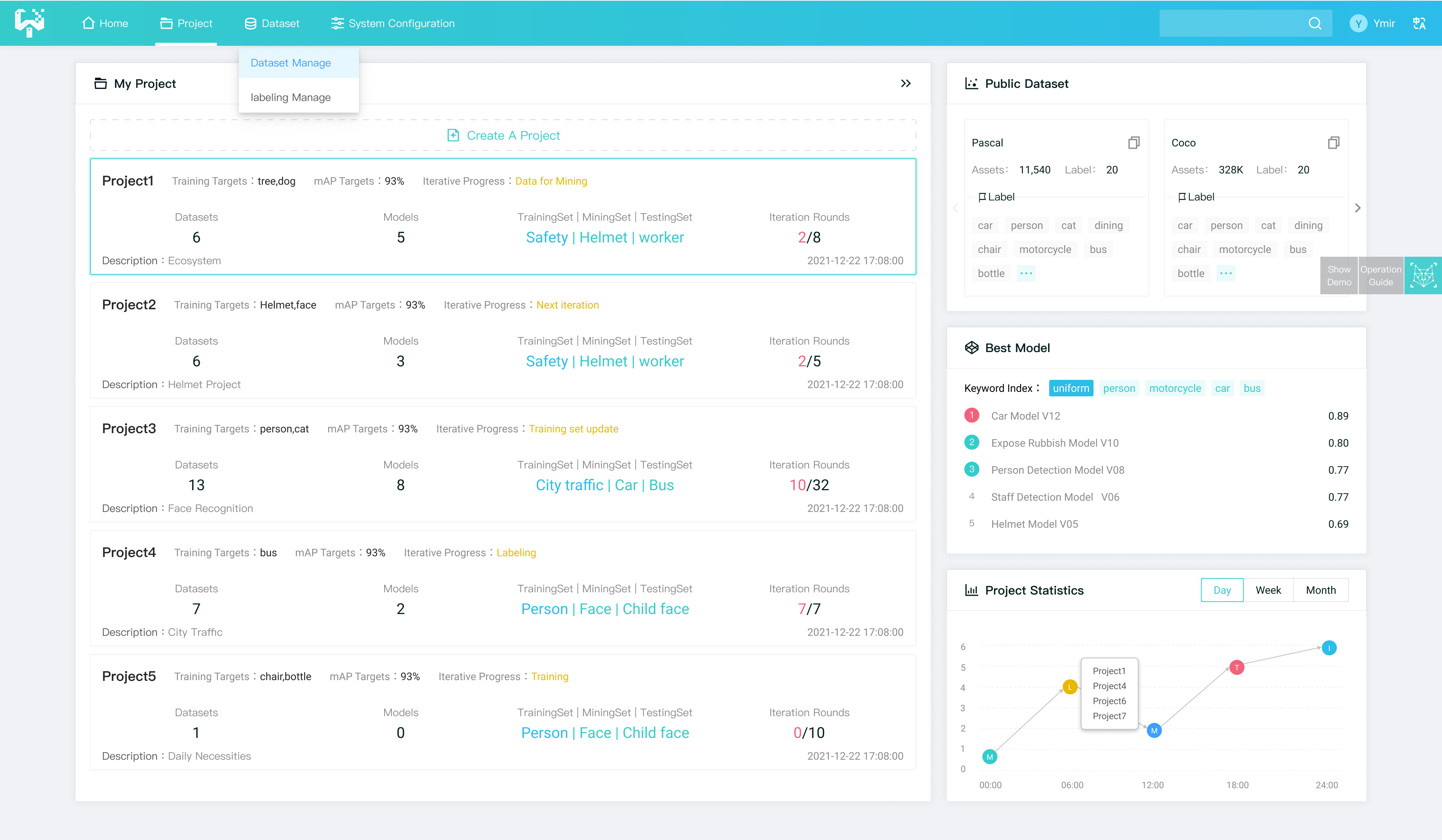}
\caption{YMIR System - Project List View}
\end{figure}

\begin{figure}[h]
\centering
\includegraphics[width=\textwidth]{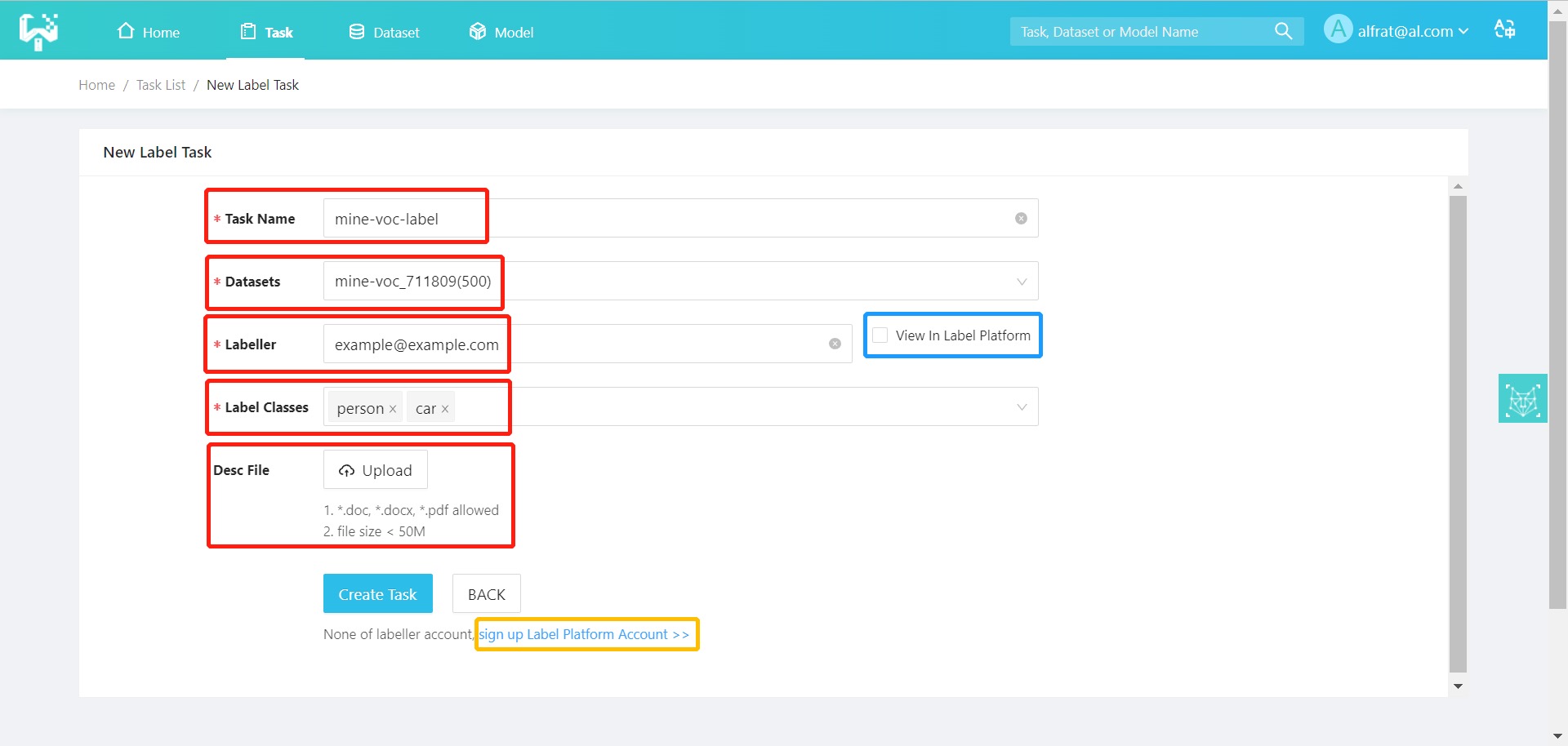}
\caption{YMIR System - Start a Labeling Task}
\end{figure}

\begin{figure}[h]
\centering
\includegraphics[width=\textwidth]{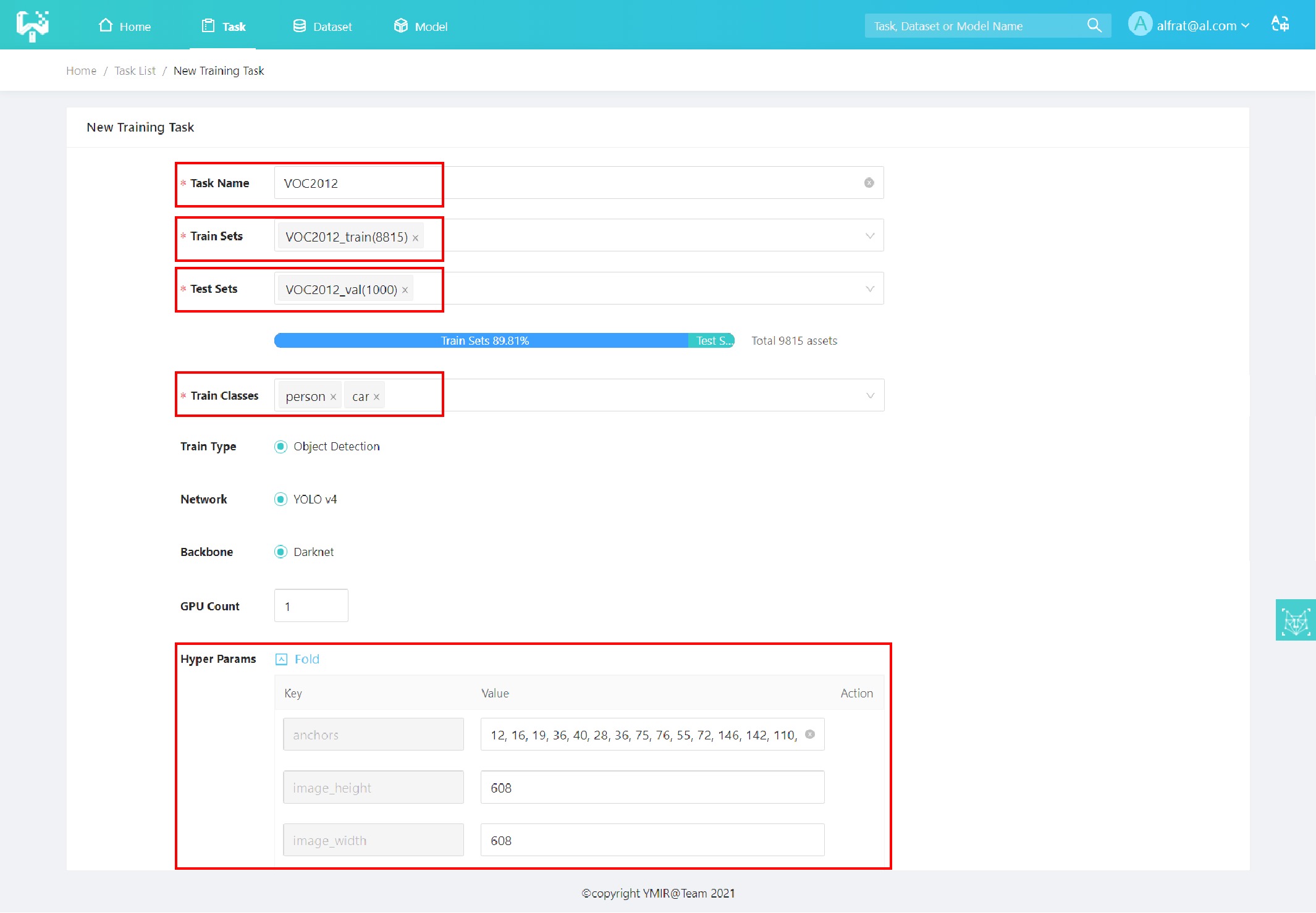}
\caption{YMIR System - Start a Training Task}
\end{figure}

\begin{figure}[h]
\centering
\includegraphics[width=\textwidth]{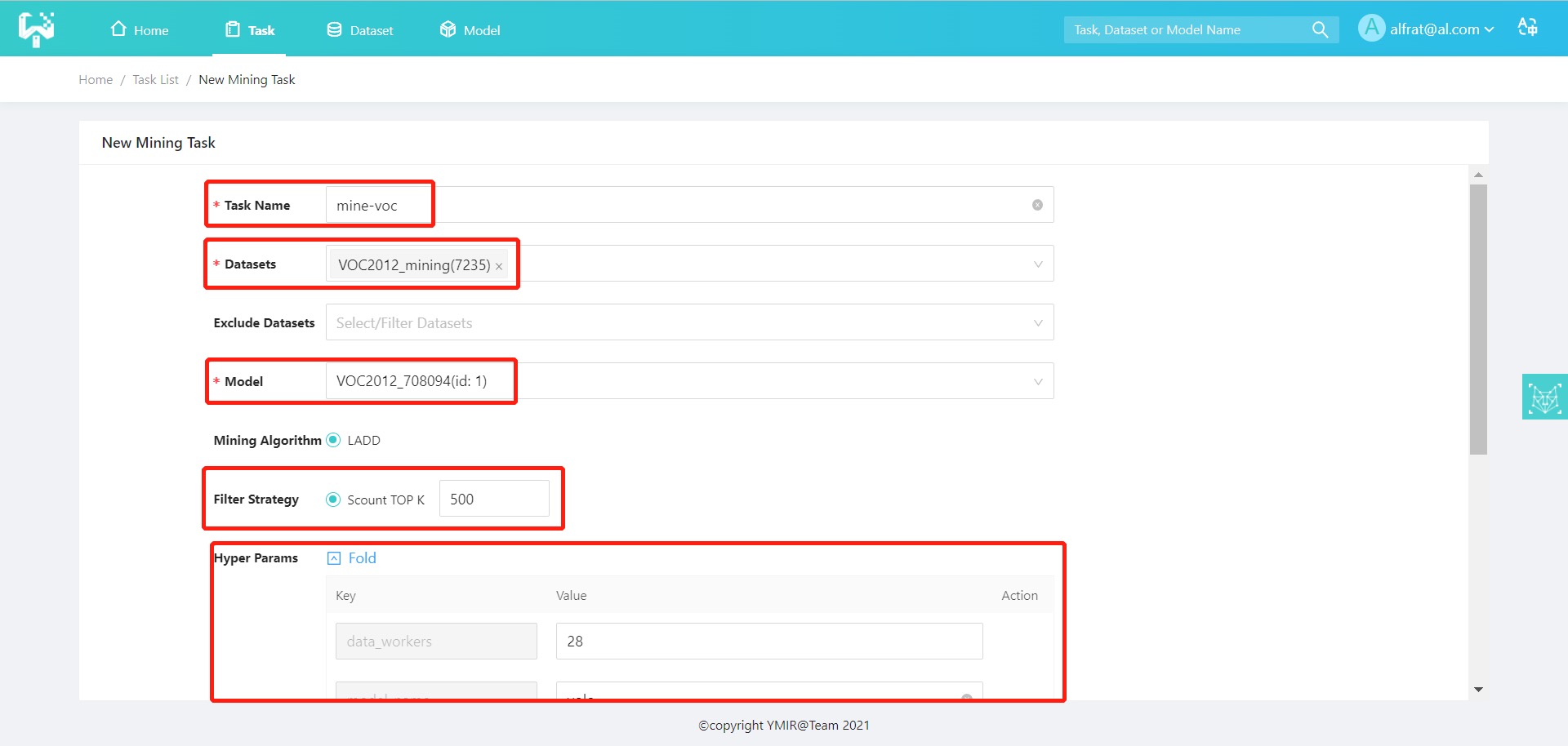}
\caption{YMIR System - Start a Data Mining Task}
\end{figure}

\begin{figure}[ht]
\centering
\includegraphics[width=\textwidth]{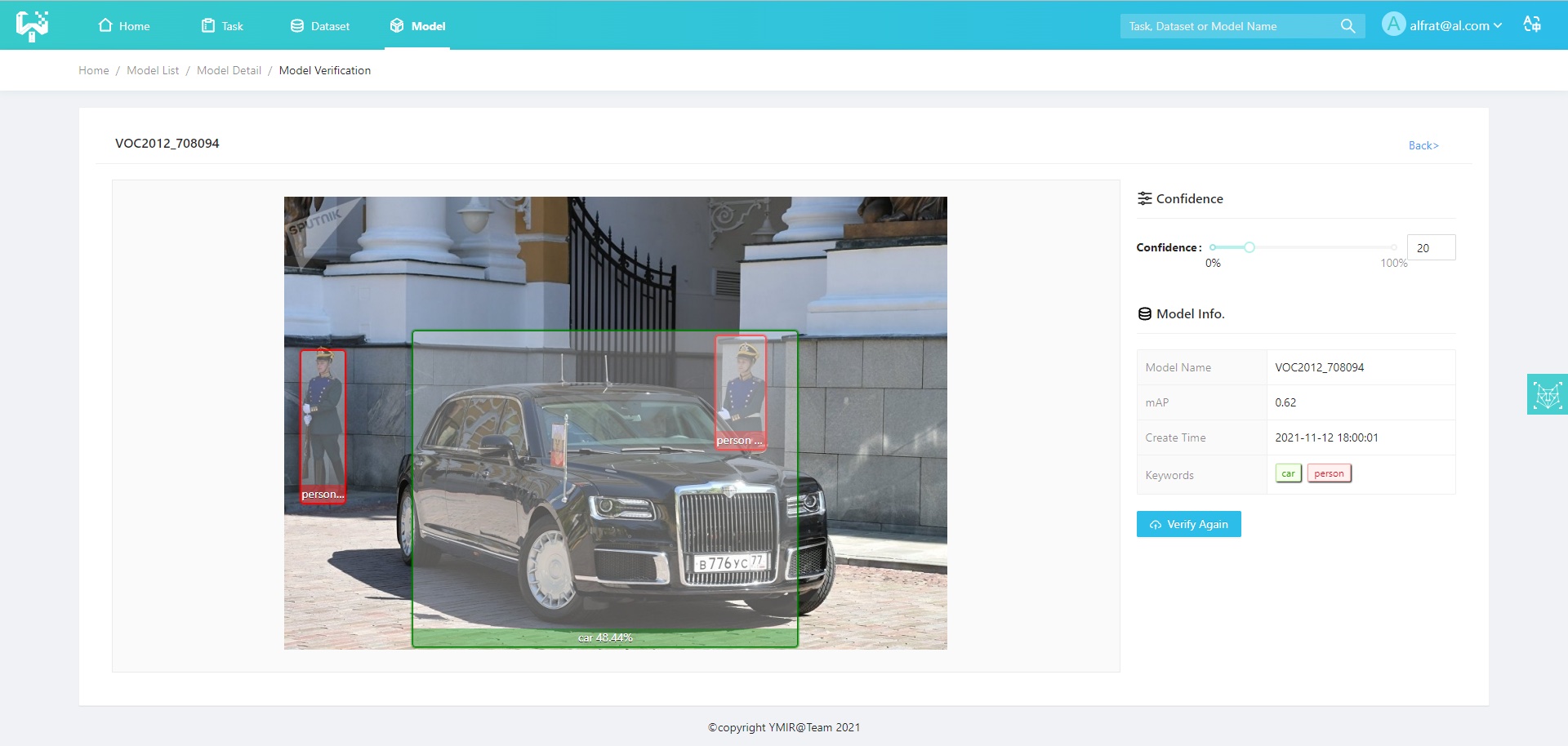}
\caption{YMIR System - Model Verification}
\end{figure}

% content will be automatically hidden during submission
%\begin{acknowledgements}
%
%\end{acknowledgements}

\end{document}